\documentclass[preprint,aps,showpacs,floatfix]{revtex4}
\usepackage{amsmath,amssymb,graphicx,epsfig}

\newcommand{\xx}{\mathbf {x}}
\newcommand{\zz}{\mathbf {z}}
\newcommand{\bs}{\mathbf {s}}

\begin{document}

\title{Reconnection of superfluid vortex bundles}

\author{Sultan Z. Alamri, Anthony J. Youd and Carlo F. Barenghi}
\affiliation {School of Mathematics and Statistics,
Newcastle University, Newcastle upon Tyne, NE1 7RU}
\date {\today}

\begin {abstract}
Using the vortex filament model and the Gross Pitaevskii
nonlinear Schroedinger equation, we show that bundles of quantised vortex 
lines in helium~II are structurally robust and
can reconnect with each other maintaining their identity. 
We discuss vortex stretching in superfluid
turbulence and show that,
during the bundle reconnection process,  
Kelvin waves of large amplitude are generated, 
in agreement with the finding that
helicity is produced by nearly singular vortex interactions
in classical Euler flows.
\end{abstract}


\maketitle

Concentrated, tubular vortical regions are commonly observed in 
turbulent flows. These vortex structures are important in 
turbulence dynamics, for example they enhance mixing and diffusion.
The role which they play in determining turbulence statistics
(spectra, intermittency) is the subject of much study\cite{Frisch}. 
If the axes of tubular vortex structures are interpreted as
the skeleton of turbulence, then the knottedness of the axes
characterises the topology, and vortex 
reconnections\cite{Kida}
are the critical events which change this topology. This
idealised picture becomes reality if one moves from ordinary fluids 
to superfluids (liquid $^4$He and $^3$He, atomic Bose--Einstein condensates
and neutron stars). The reason is that in superfluids
quantum mechanics constrains any rotational motion to ultra thin
vortex filaments whose core radius and circulation are fixed, and
turbulence takes the form of a tangle of such
discrete filaments \cite{Donnelly}. 
Recent experimental, theoretical and numerical studies have
revealed similarities between ordinary turbulence and 
superfluid turbulence \cite{similarities}, 
such as the same Kolmogorov -5/3 energy spectrum
\cite{spectrum}
and the same temporal decay law of the vorticity\cite{decay}.

This letter is concerned with vortex reconnections and vortex
stretching in superfluid turbulence. The importance in helium~II of
vortex reconnections and its scaling laws\cite{CFB-scaling}
was first appreciated by Schwarz\cite{Schwarz}, who developed
the reconnecting vortex filament model; later, the 
existence of reconnections was
confirmed by Koplik and Levine\cite{Koplik}
using the nonlinear Schroedinger equation (NLSE) model.  
Vortex reconnections are associated with the
dissipation of superfluid kinetic energy in the limit of absolute
zero\cite{Vinen-sound}, either directly\cite{pulse} or via a Kelvin wave 
cascade\cite{Kelvin-cascade}. Recently, individual vortex
reconnections have been detected in experiments\cite{Sreeni}.
As recognised by
Procaccia and Sreenivasan\cite{Procaccia-Sreeni}, the 
occurrence of the -5/3 law in superfluid turbulence is surprising
if one notices that vortex stretching, usually recognised as an
important mechanism to transfer energy across length scales, is absent
because the radius of the superfluid vortex core is fixed. A possible
solution of the puzzle is that some filaments
are organised in vortex bundles, which have indeed been
noticed in the most recent numerical simulations of
superfluid turbulence \cite{bundles}.
Clearly, stretching occurs if the relative position of vortex
strands within a bundle changes during the evolution. 


In a first set of numerical experiment we use the
model of Schwarz\cite{Schwarz}. Let the space curve ${\bs}(\xi,t)$
represent a vortex filament where $\xi$ is arclength and $t$ is
time. The self--induced velocity ${\bf v}_{si}$ of the vortex at ${\bf s}$ is 

\begin{equation}
{\bf v}_{si}=\frac{d\bf s}{dt}=-\frac{\kappa}{4 \pi} \oint \frac{(\xx-\zz) }
{\vert \xx - \zz \vert^3}
\times {\bf d}\zz,
\label{eq:BS}
\end{equation}

\noindent
where $\kappa$ is the quantum of circulation 
and the Biot--Savart integral is suitably desingularized\cite{Schwarz}. The 
numerical method to evolve configurations of vortex filaments
has been described elsewhere \cite{Schwarz,ABC}.
Here it suffices to say that the filaments are
discretised into a large, variable number of points, $N$,
depending on the 
local curvature: points are removed in regions where 
filaments straighten and are added where the local 
radius of curvature becomes smaller. The time evolution is based on
a $4^{th}$ order Runge--Kutta scheme with variable time step
$\Delta t$, which depends
on the current minimum distance $\delta_{min}$ along points, which
determines the frequency of rotation of the fastest Kelvin wave
perturbation. 
If two vortex strands become closer to each other than the local
discretization along filaments, then, consistently with the orientation
of the filaments, our numerical algorithm reconnects the strands
\cite{Tsubota}, provided that the total length is decreased\cite{Samuels}.
This condition\cite{proxy} mimics the small kinetic energy losses at
vortex reconnections discussed in Ref.~\cite{pulse}. We also
checked that, with our algorithm, the reconnection condition used by
Schwarz\cite{Schwarz} is satisfied\cite{note}.

\begin{figure}[ht]
\begin{minipage}[b]{0.4\linewidth}
\centering
\includegraphics[angle=-90,scale=0.35]{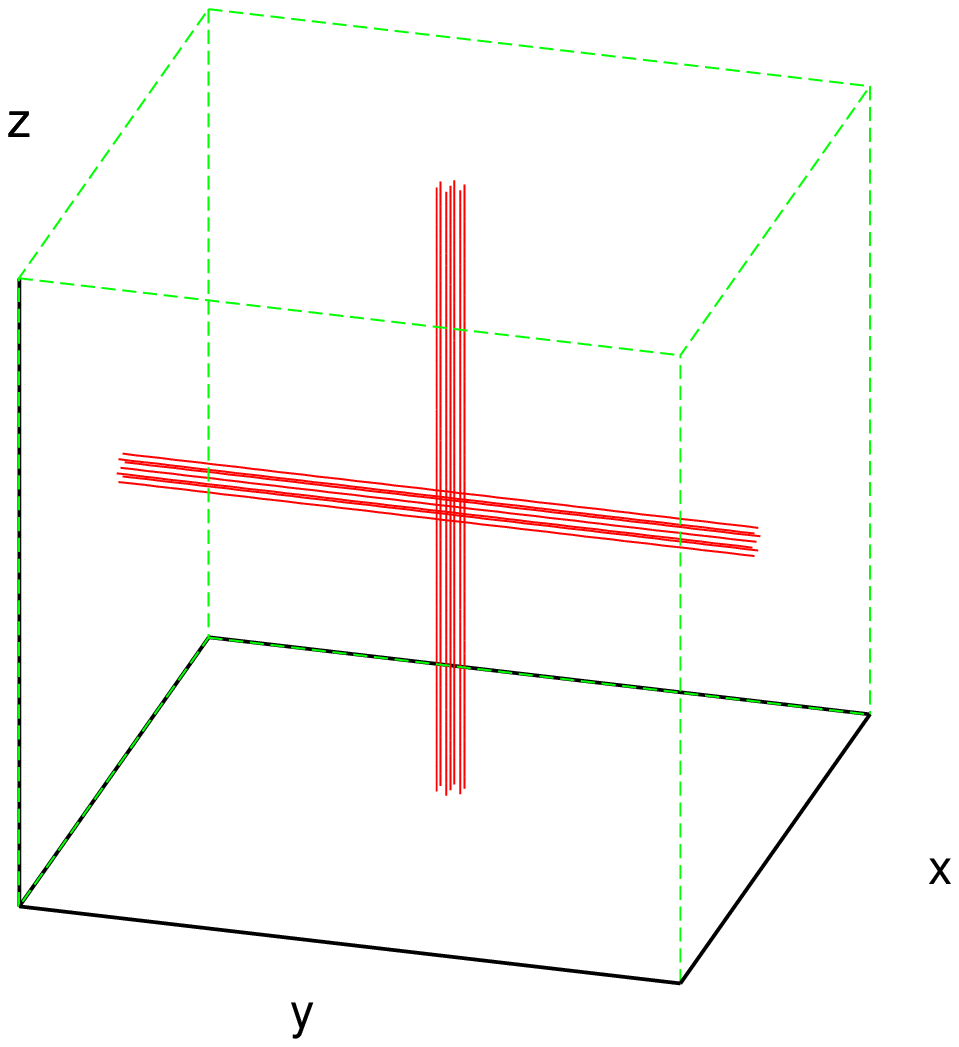}
\end{minipage}
\hspace{0.5cm}
\begin{minipage}[b]{0.4\linewidth}
\centering
\includegraphics[angle=-90,scale=0.35]{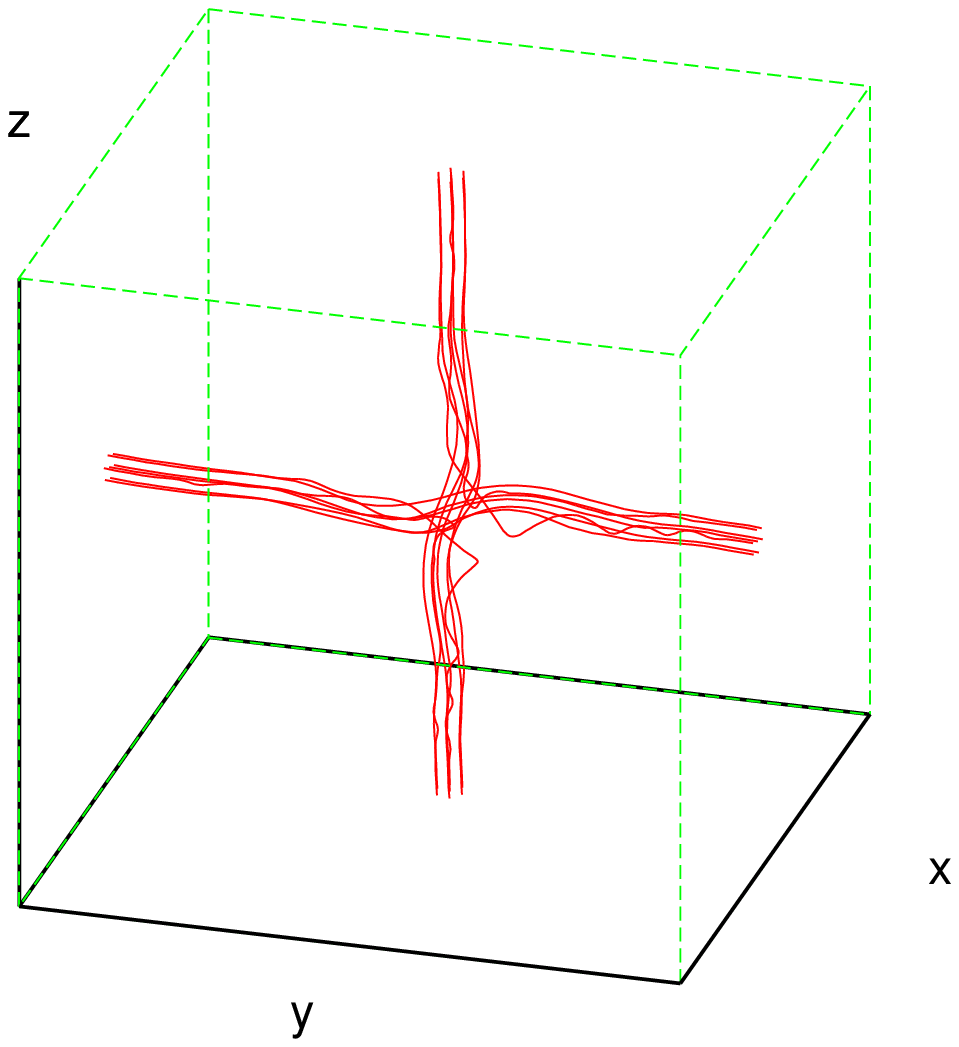}
\end{minipage}

\vspace{0.5cm}

\begin{minipage}[b]{0.4\linewidth}
\centering
\includegraphics[angle=-90,scale=0.35]{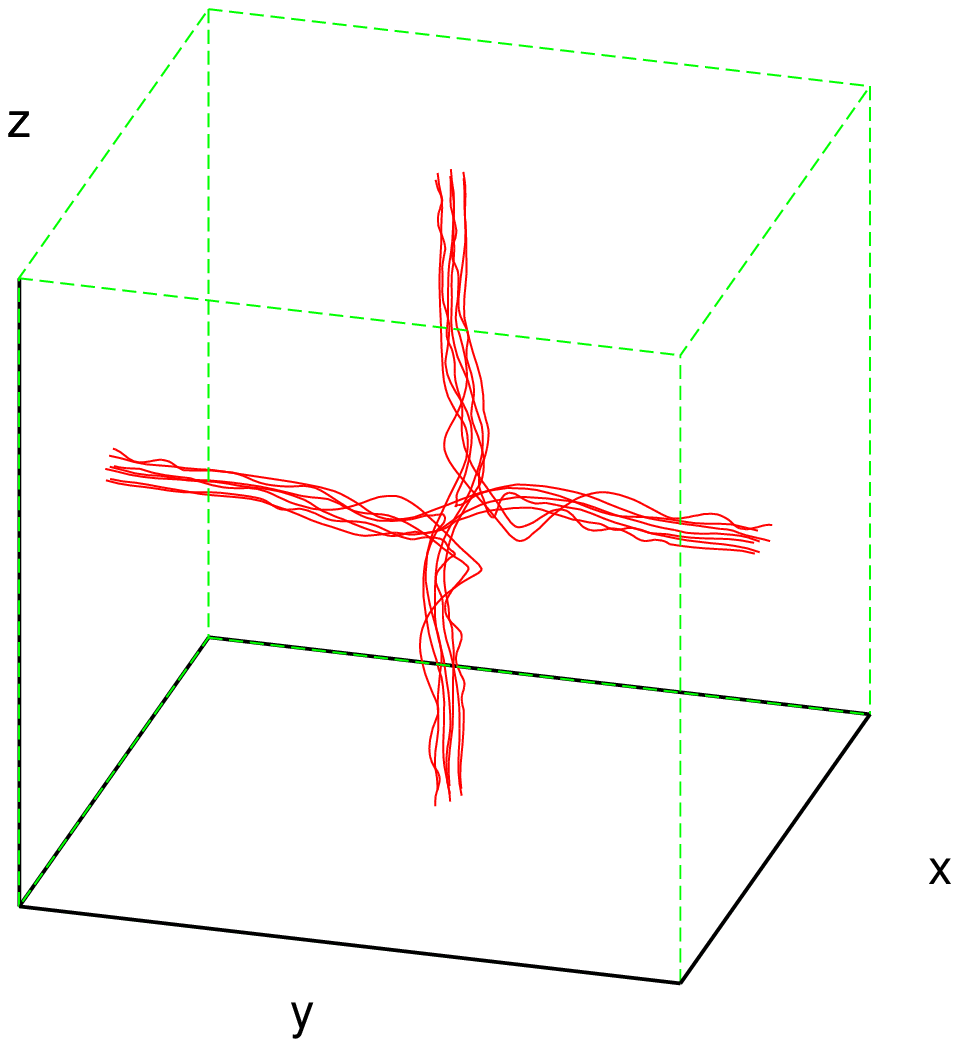}
\end{minipage}
\hspace{0.5cm}
\begin{minipage}[b]{0.4\linewidth}
\centering
\includegraphics[angle=-90,scale=0.35]{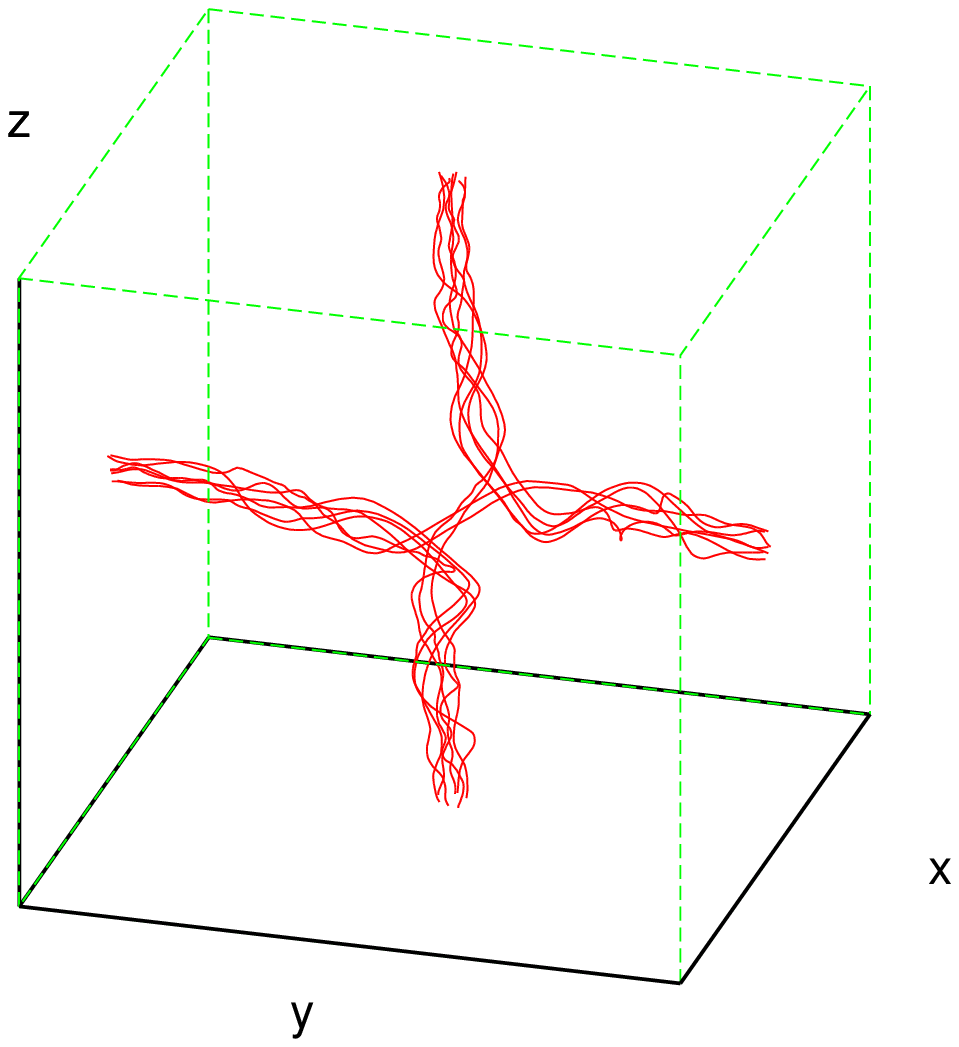}
\end{minipage}

\vspace{0.5cm}

\begin{minipage}[b]{0.4\linewidth}
\centering
\includegraphics[angle=-90,scale=0.35]{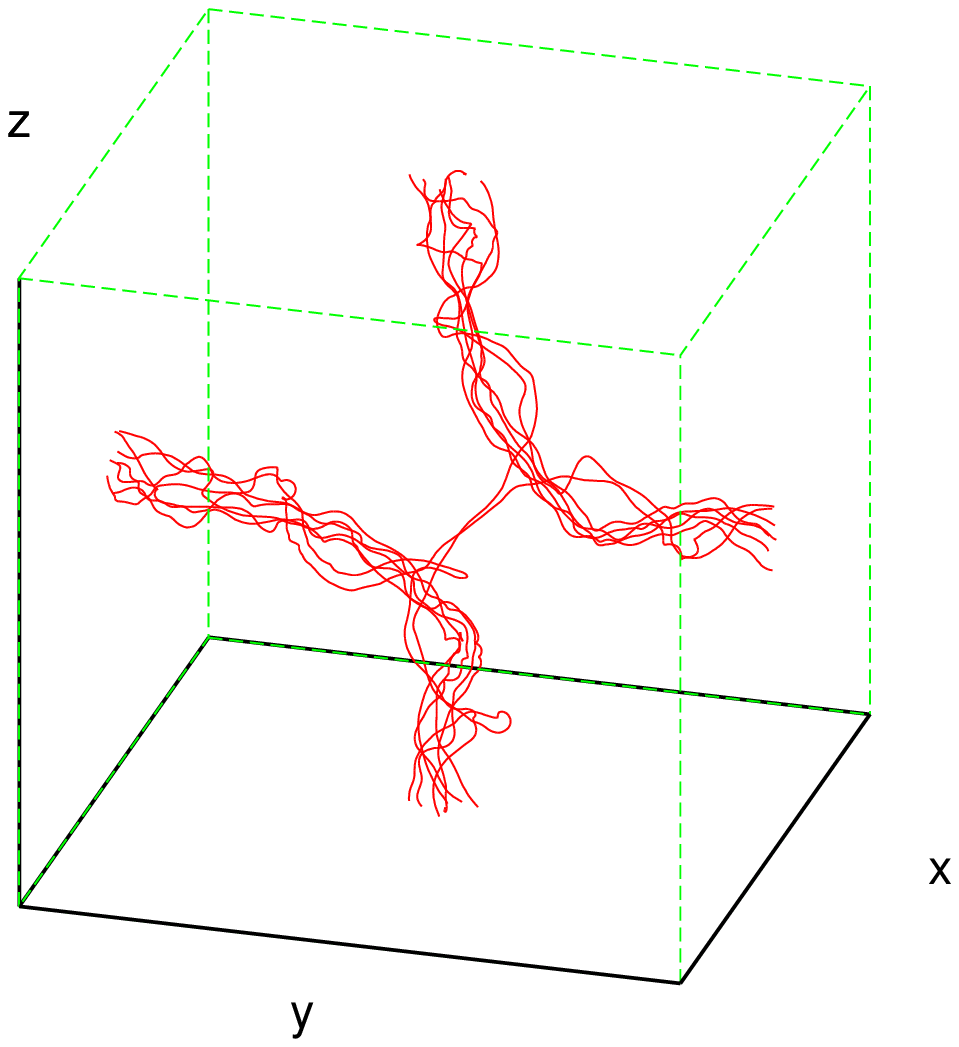}
\end{minipage}
\hspace{0.5cm}
\begin{minipage}[b]{0.4\linewidth}
\centering
\includegraphics[angle=-90,scale=0.35]{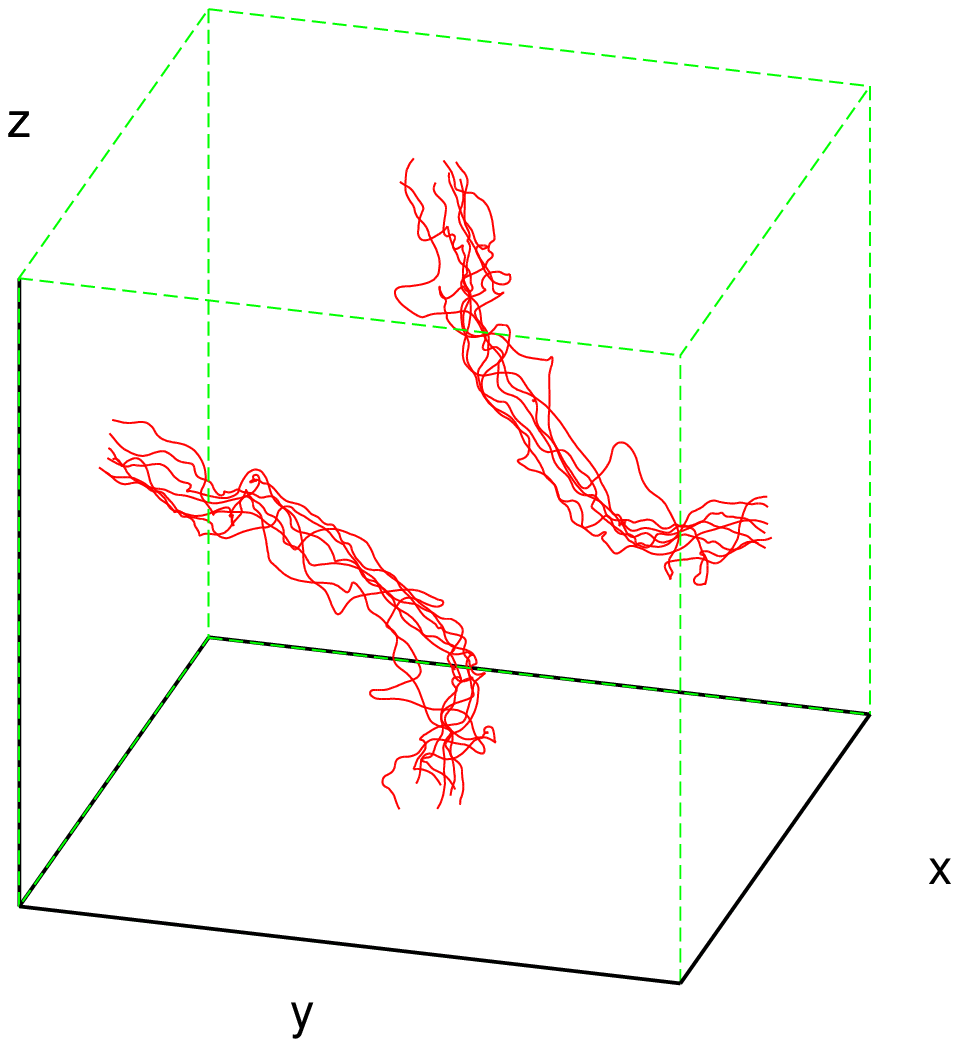}
\end{minipage}

\caption{(Colour online).
Reconnection of two bundles of seven
vortex strands each. 
a, top left: $t=0~\rm s$.
b, top right: $t=7.13~\rm s$.
c, middle left: $t=23.58~\rm s$.
c, middle left: $t=36.27~\rm s$.
d, bottom left: $t=61.49~\rm s$.
e, bottom right: $t=80.35~\rm s$.
}
\label{fig1}
\end{figure}

Using this model, we study the interaction of two bundles of a given
number $M$ of (initially) straight parallel vortex strands, 
set (initially) at 90 degrees to each other. 
Le $D$ be the distance
between the axes of the two bundles and $A$ be the radius of each bundle.
The calculation is performed in a cubic periodic box $-B \leq x,y,z \leq B$.
A typical result for $M=7$, 
$A=0.0417~\rm cm$, $D=3A$ and $B=1~\rm cm$ is shown in Fig.~\ref{fig1}.
The initial position of vortex strands within the same bundle is symmetric:
in this case we place six vortices at the corner of a hexagon and one
vortex in the middle.
In the absence of the the second bundle, the vortex strands of
the first bundle would rotate around each other in a time scale
$\tau$ of the order of $10~\rm s$ because they are parallel. 
With two bundles, the interaction makes them 
to bend in the direction of each other, until, at time 
$t_1\approx 6.7~\rm s$, the first reconnection 
occurs. The successive evolution involves the reconnections 
of all strands at times
$t_2\approx 16.4~\rm s$,
$t_3\approx 21.0~\rm s$,
$t_4\approx 23.6~\rm s$,
$t_5\approx 27.4~\rm s$,
$t_6\approx 52.0~\rm s$,
and $t_7\approx 65.9~\rm s$,
after which the two bundles separate from each other 
(Fig.~\ref{fig1}, bottom right) and move away. 

\begin{figure}[!h]
\includegraphics[angle=-90,width=0.4\textwidth]{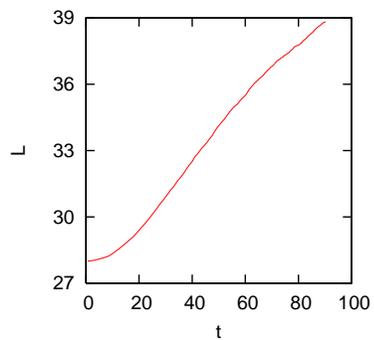}
\caption{(Colour online). 
Total vortex length $L$ vs time $t$ corresponding to 
Fig.~\ref{fig1}.
}
\label{fig2}
\end{figure}

\begin{figure}[!h]
\includegraphics[angle=-90,width=0.4\textwidth]{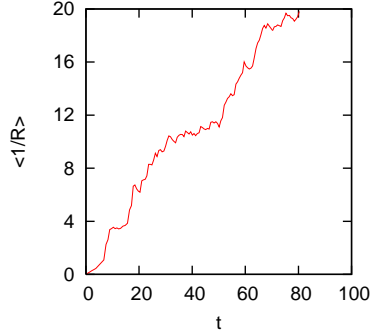}
\caption{(Colour online).
Average inverse radius of curvature $<1/R>$ vs time $t$
corresponding to Fig.~\ref{fig1}.
}
\label{fig3}
\end{figure}

\begin{figure}[!h]
\includegraphics[angle=-90,width=0.4\textwidth]{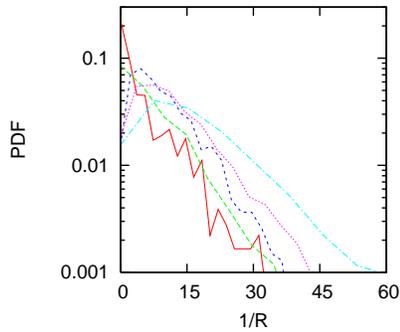}
\caption{(Colour online).
PDF of the inverse radius of curvature at different
times, corresponding to  
Fig.~\ref{fig1}. From left to right, the times are:
(solid red): $t=7.13~\rm s$,
(solid green) $t=23.58~\rm s$,
(dashed blue) $t=36.27~\rm s$,
(solid purple) $t=61.49~\rm s$,
(solid light blue) $t=80.35~\rm s$.
}
\label{fig4}
\end{figure}

Similar calculations with different parameters show that
the last reconnection (at which the remaining vortex bridge breaks, and
the bundles become free from
each other) usually takes longer than the first few reconnections.
Calculations also show that, after the first
reconnection, the vortex strands develop Kelvin waves of large
amplitude.
During the process, the total vortex length $L$
increases by about $40 \%$, as shown in Fig.~\ref{fig2}.
Fig.~\ref{fig3} shows the average inverse
radius of curvature, $<1/R>$, obtained by computing 
$\vert d^2{\bf s}/d\xi^2 \vert$
at each point  ${\bf s}_j$ ($j=1,\cdots N$)
and then averaging over all points. 
To cope with the increase in $L$ and decrease of $<R>$,
the number of points (initially $N=700$) grows with time, up to $N=4191$,
when we stop this particular calculation.
Fig.~\ref{fig4} shows 
normalised histograms (Probability Density Functions) of the average
inverse curvature computed at different times: the formation of
the Kelvin wave
cascade is evident in the development of a tail of the distribution
at larger and larger times. All these results were confirmed
by further calculations with $M=5$ and $11$ strands.

To make sure that our result does not depend on the 
reconnection algorithm, we study the interaction of
vortex bundles by solving the 
NLSE, also called the Gross--Pitaevksii (GP) equation:

\begin{equation}
2 i \frac{\partial \psi}{\partial t}
=-\nabla^2 \psi + \vert \psi \vert^2 \psi -\psi.
\label{eq:nlse}
\end{equation}

\noindent
The NLSE is a convenient model of superfluid 
helium~II \cite{Roberts-Berloff}; for example, Koplik
and Levine \cite{Koplik} used it to confirm
Schwarz's insight that quantised vortices reconnect.
Eq.~\ref{eq:nlse} is written in terms of
$\hbar /\sqrt{2mE}$ (the coherence length) as
unit of space,
$2E/\hbar$ as unit of time and $mE/V$ as unit of density, where $m$ is the
mass of one boson, $V$ the strength of the interaction between bosons,
and $E$ the energy per boson. 
The calculation is done using a $5^{th}$ order Runge Kutta Fehlberg
method (with typical time step of the order of $0.01$) 
using $256^3$ grid points and reflective boundary conditions.

\begin{figure}[ht]
\begin{minipage}[b]{0.4\linewidth}
\centering
\includegraphics[angle=0,scale=0.20]{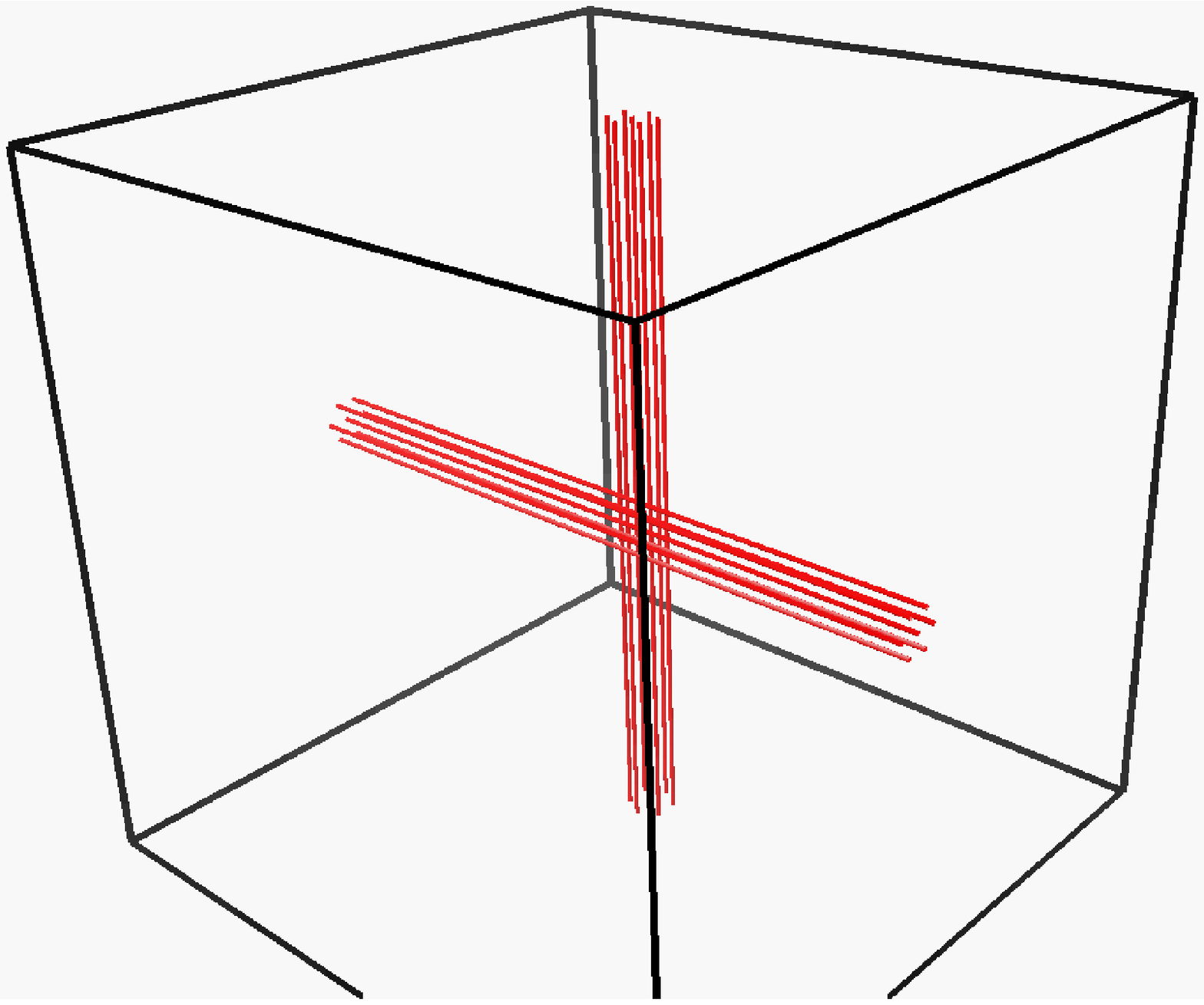}
\end{minipage}
\hspace{0.5cm}
\begin{minipage}[b]{0.4\linewidth}
\centering
\includegraphics[angle=0,scale=0.20]{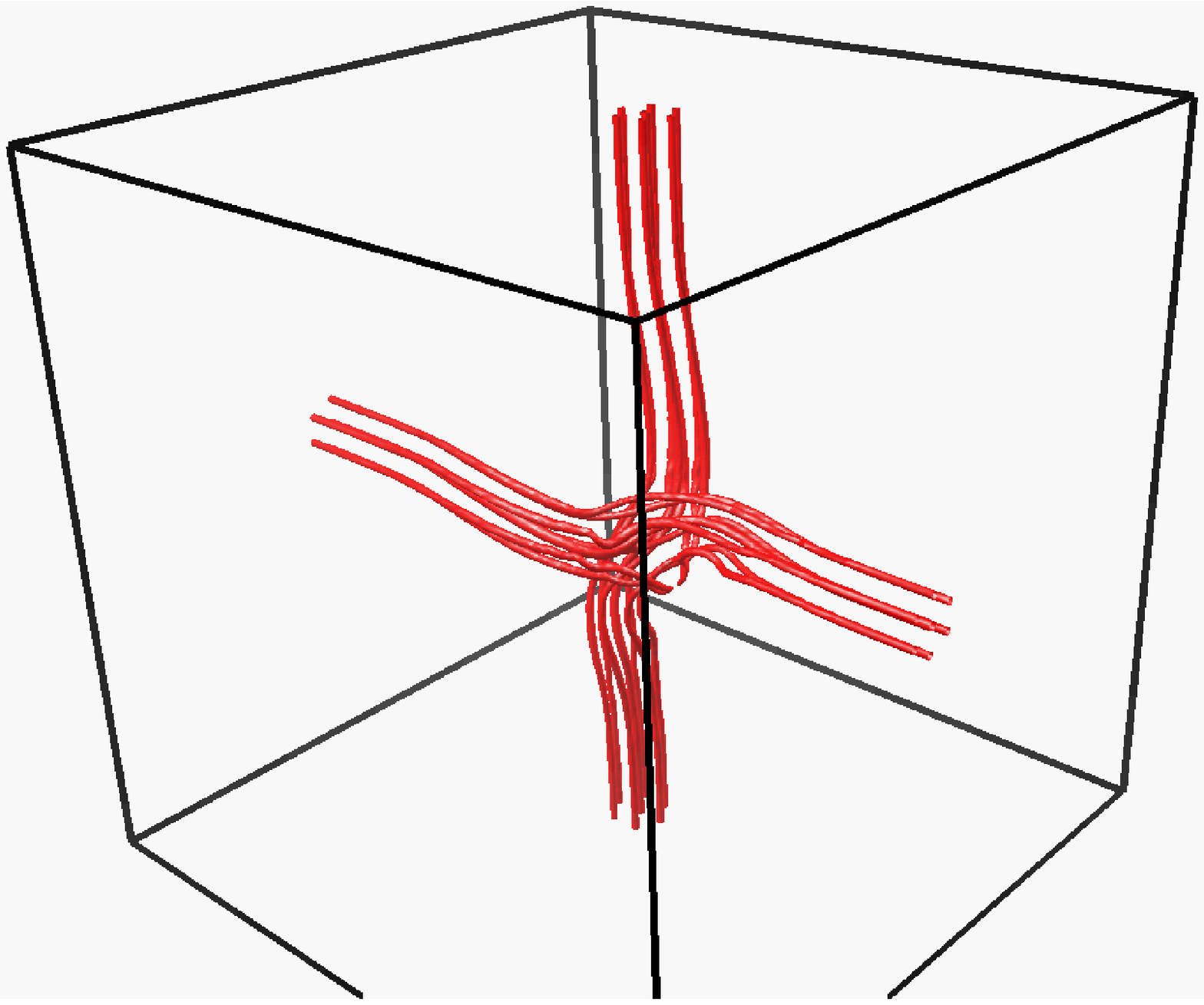}
\end{minipage}

\vspace{0.3cm}

\begin{minipage}[b]{0.4\linewidth}
\centering
\includegraphics[angle=0,scale=0.20]{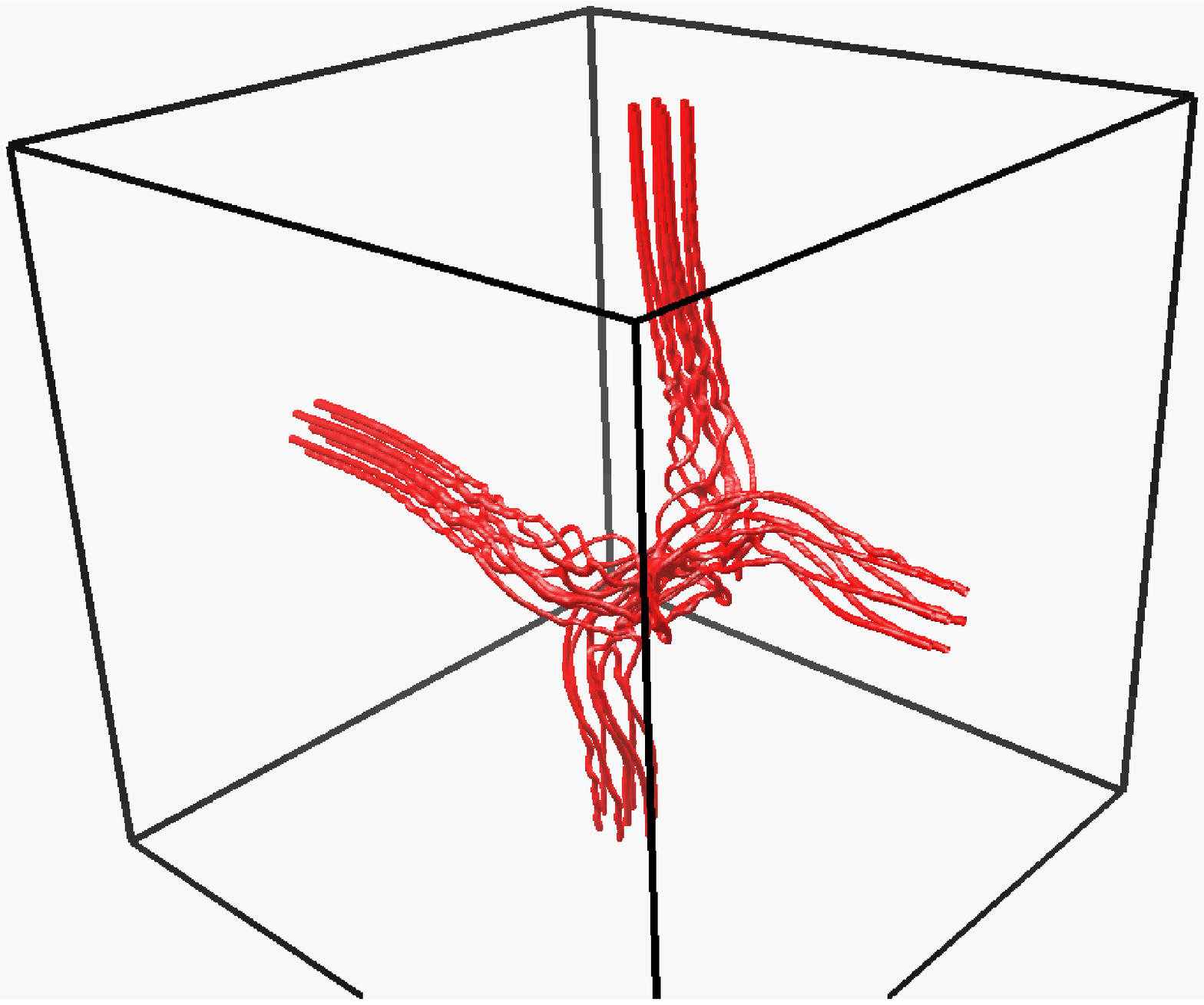}
\end{minipage}
\hspace{0.5cm}
\begin{minipage}[b]{0.4\linewidth}
\centering
\includegraphics[angle=0,scale=0.20]{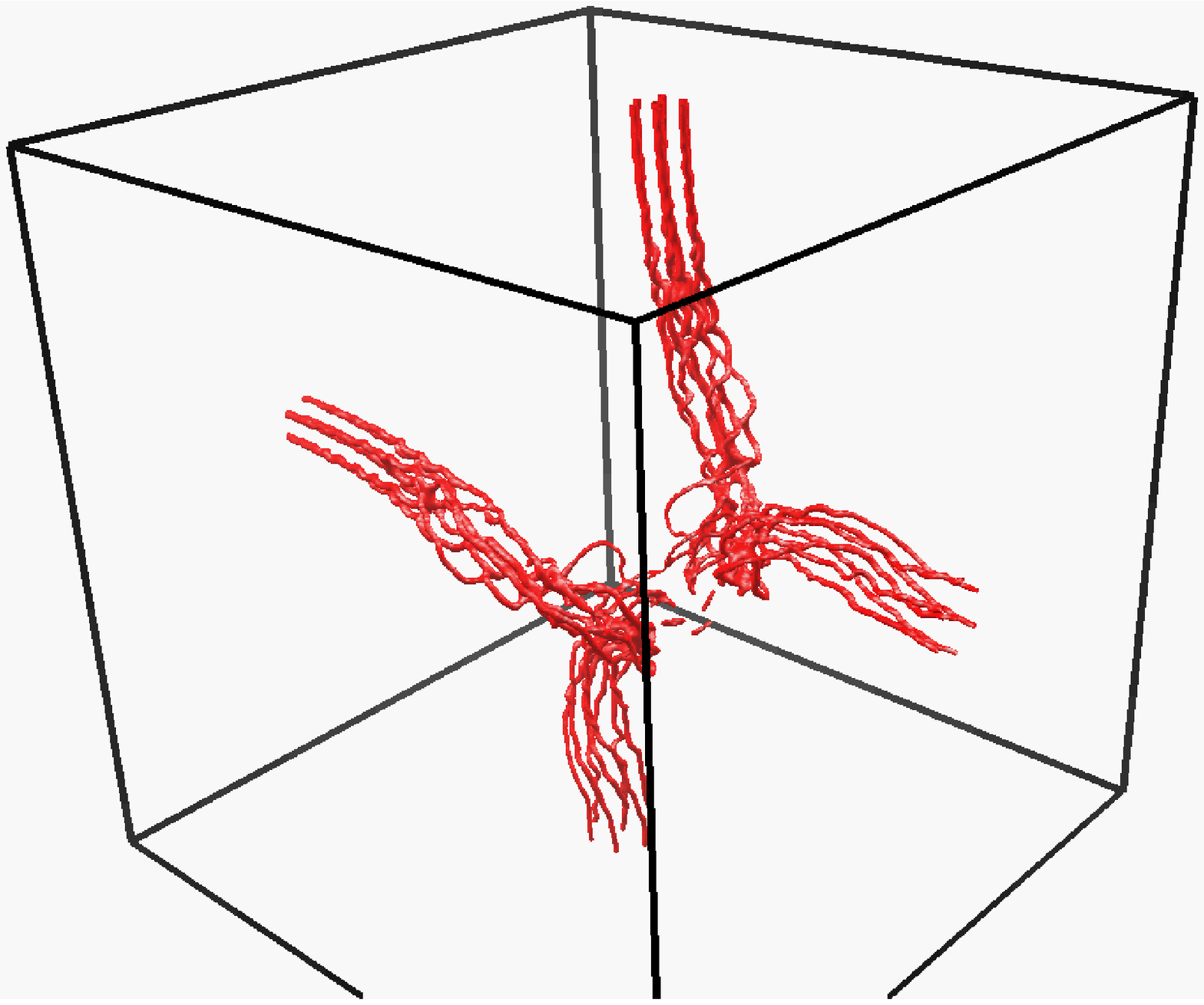}
\end{minipage}

\vspace{0.3cm}

\begin{minipage}[b]{0.4\linewidth}
\centering
\includegraphics[angle=0,scale=0.20]{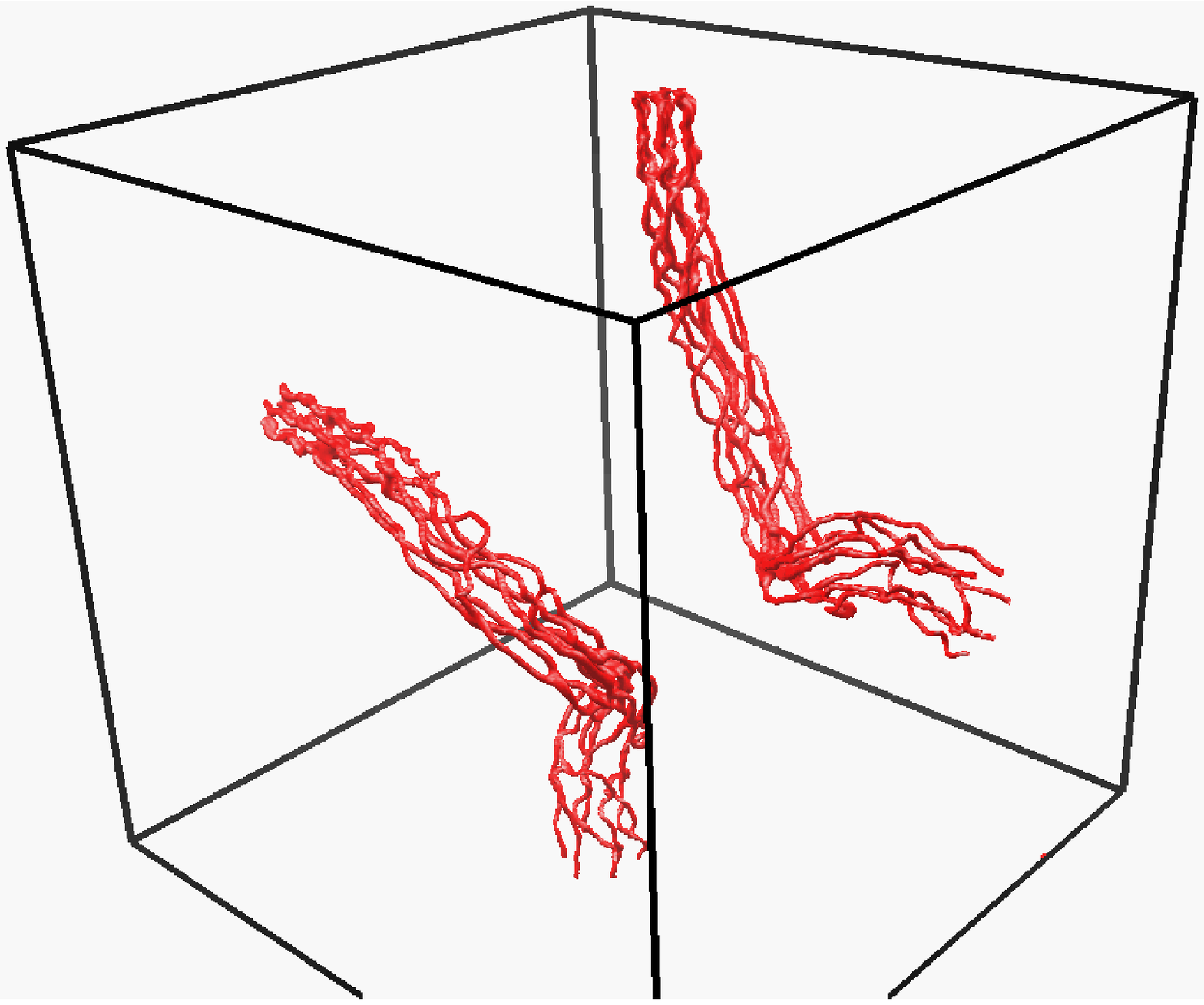}
\end{minipage}
\hspace{0.5cm}
\begin{minipage}[b]{0.4\linewidth}
\centering
\includegraphics[angle=0,scale=0.20]{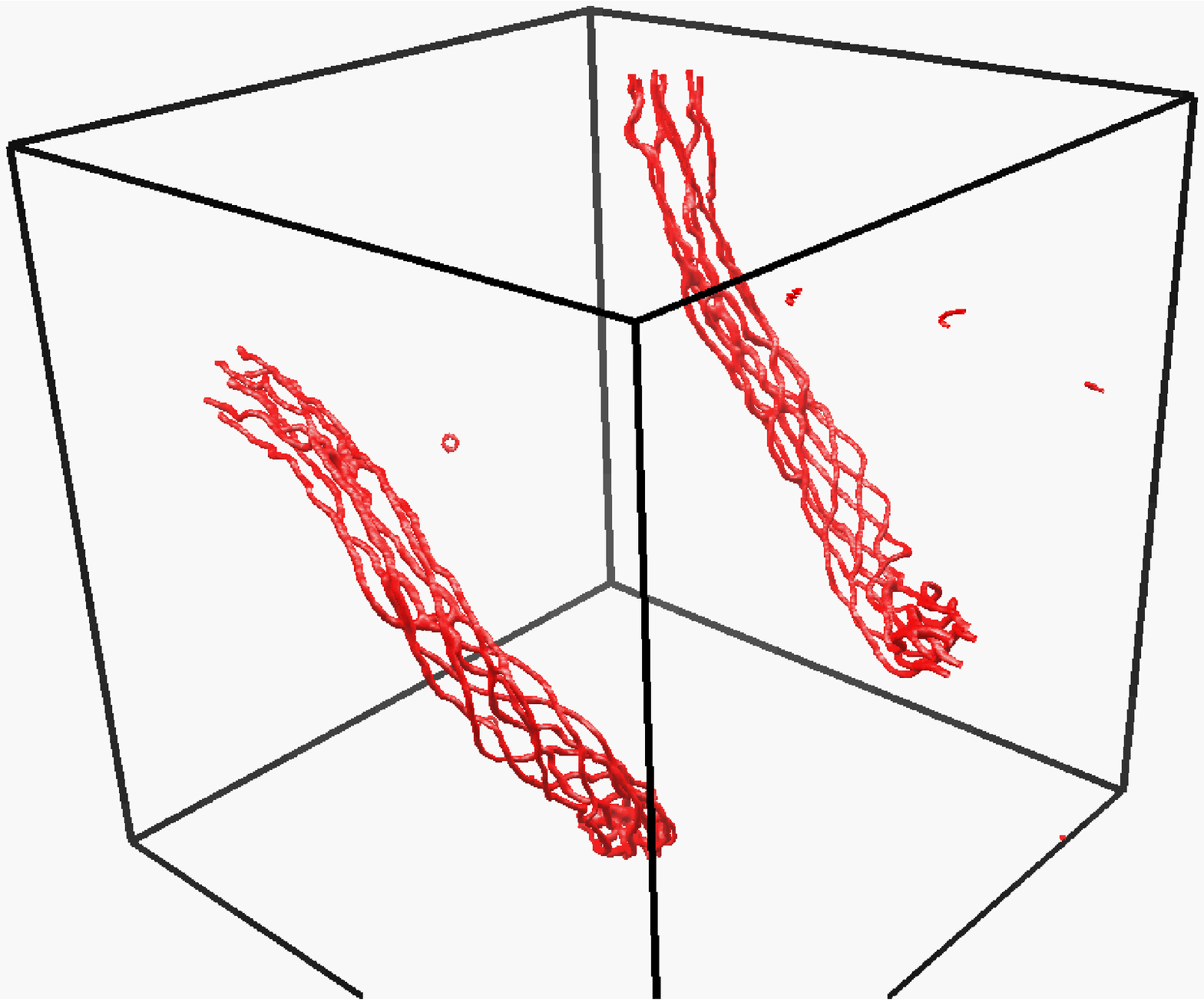}
\end{minipage}

\vspace{0.3cm}
\caption{
Reconnection of two bundles of
seven vortices each computed by solving the dimensionless
NLSE.  The figures show density isosurfaces 
at the level $0.25$ (of the unit bulk density away from vortices) at
different times $t$. 
The numerical resolution of the vortex core is such that at $0.25$ density
level there are about $3$ grid points within a vortex core, and at $0.90$
density level there are about $15$ grid points.
Top right: $t=110$;
Middle left: $t=240$;
Middle right: $t=320$;
Bottom left: $t=450$;
Bottom right: $t=800$.}
\label{fig5}
\end{figure}

\begin{figure}[!h]
\includegraphics[angle=0,width=0.40\textwidth]{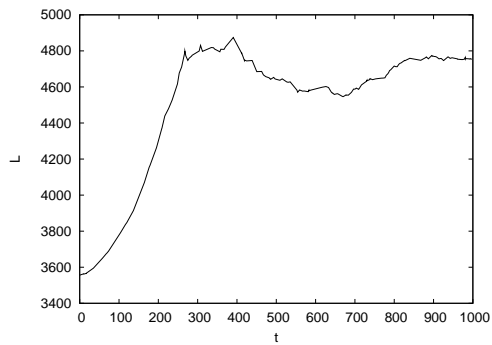}
\caption{
Dimensionless vortex length $L$ vs dimensionless
time $t$ corresponding to Fig.~\ref{fig5}.}
\label{fig6}
\end{figure}

In a second set of calculations, we thus
solve the NLSE in a cubic box $-128 \leq x,y,z \leq 128$.
The initial condition consists of two bundles of radius $A=8$ and distance
$D=3A$ between the axes of the bundles. Each bundle contains $M=7$ vortices
(six at the corners of a hexagon and one in the middle). 
The time sequence, shown in Fig.~\ref{fig5}, confirms the previous
result that vortex bundles maintain their identity
and reconnect with each other.  Note the emission of small
vortex rings in the last image.  As in the previous
calculation, large amplitude Kelvin waves are
generated. 
Fig.~\ref{fig6} shows that at first $L$ increases with $t$, then,
when the bundles have moved sufficiently away from each other,
$L$ saturates; the relative increase of $L$
is about $30 \%$, confirming the intensification
of vorticity.  The relation
between energy and length is important; it is common
in the literature to state that the energy per unit length of
vortex line is
$(\rho_s \kappa^2/(4 \pi)) \ln{(b/a})$ where $\rho_s$ is the 
superfluid density and $b$ is an upper cutoff (the radius of the
container or the distance to the next vortex), but this relation
is valid only for a straight vortex of course. The NLSE
model allows to determine the energy more precisely than the
vortex filament model: in our calculation the total
mass and the total energy  are conserved within $1$ part
in $10^3$ and $1$ part in $10^4$ respectively. Further calculations
with $M=5$ and $9$ strands confirm our results.

\begin{figure}[ht]
\begin{minipage}[b]{0.4\linewidth}
\centering
\includegraphics[angle=-90,scale=0.35]{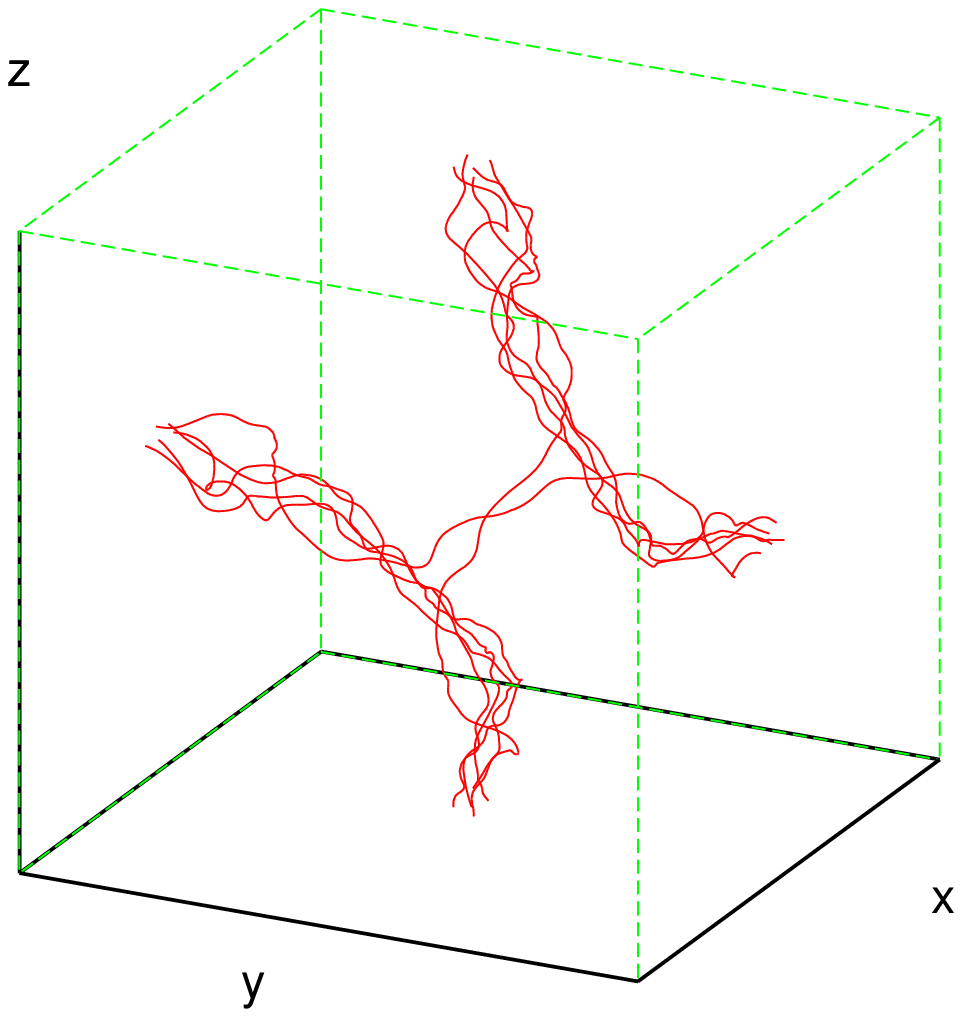}
\end{minipage}
\hspace{0.5cm}
\begin{minipage}[b]{0.4\linewidth}
\centering
\includegraphics[angle=-90,scale=0.35]{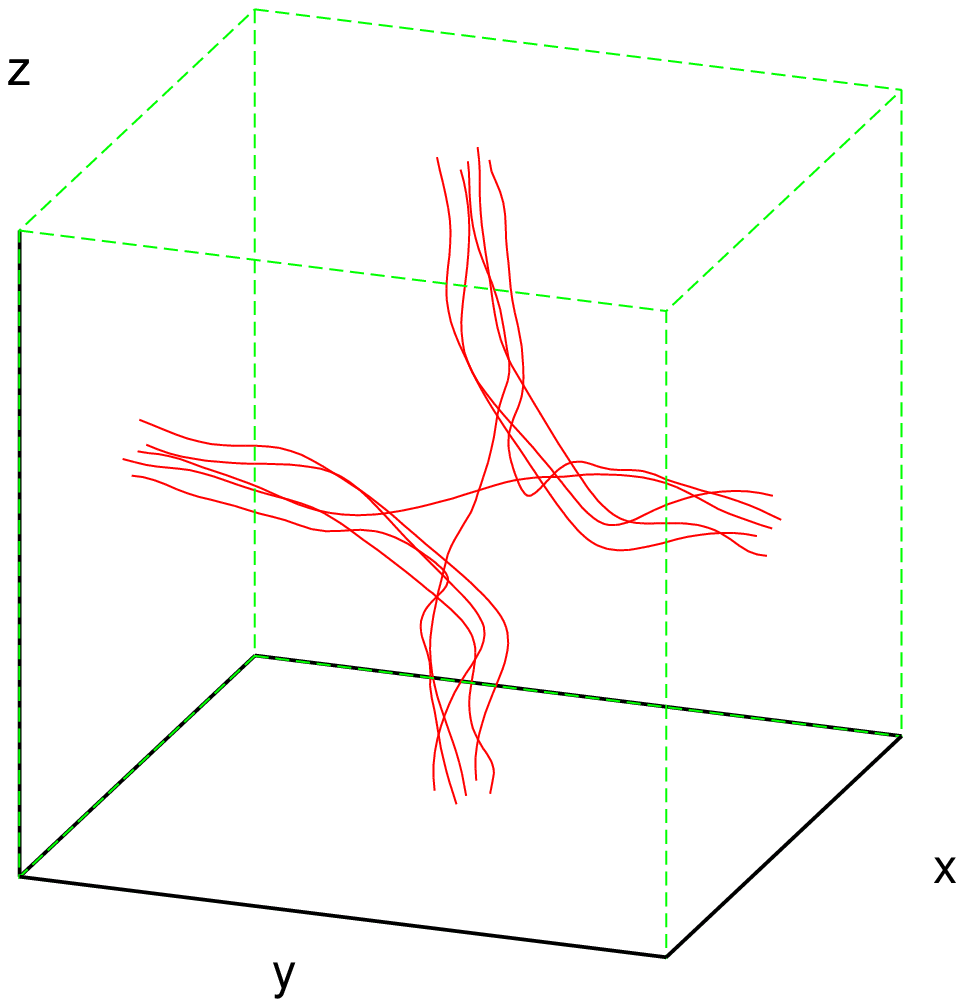}
\end{minipage}

\caption{Comparison between bundles reconnections at
$T=0~\rm K$ (left, corresponding to
$\alpha=\alpha'=0$) and at $T=1.65~\rm K$ (right, corresponding
to $\alpha=0.111$ and $\alpha'=0.01437$). The initial condition
is the same for both calculations. The times (left:
$t=86.9~\rm s$, right: $t=58.1~\rm s$) are chosen so that in each
case the bundle reconnection has proceed to the point that only
two vortex strands are still part of the initial bundle.
}
\label{fig7}
\end{figure}

The above results refer to temperature $T=0$.
If $T>0$ the motion of a vortex filament is 
governed by Schwarz's equation

\begin{equation}
\frac{d{\bf s}}{dt}={\bf v}_{si}
+ \alpha {\bf s}' \times ({\bf v}_n -{\bf v}_{si})
- \alpha' {\bf s}' \times [{\bf s}' \times ({\bf v}_n -{\bf v}_{si})],
\label{eq:schwarz}
\end{equation}

\noindent
where ${\bf s}'=d{\bf s}/d\xi$ is the unit tangent at ${\bf s}$, ${\bf v}_n$
is the normal fluid velocity and $\alpha$ and
$\alpha'$ are known temperature dependent coefficients resulting
from the mutual friction between the vortex lines and the 
normal fluid \cite{BDV}.  In a third set of calculations we find
that, at $T=1.65~\rm K$ (the typical temperature of many experiments),
with ${\bf v}_n=0$ , 
bundle reconnections are still possible, although with much less
helical disturbances, as shown in Fig.~\ref{fig7}, in agreement 
with calculations \cite{Tsubota}
which show that at $T>0$ the vortex tangle is 
much smoother than at $T=0$. We do not repeat this calculation
with the NLSE model because there is not yet a consensus on how
to generalise the NLSE to finite temperatures, although many approaches have
been proposed \cite{finitet}.

We conclude that vortex bundles are structurally stable structures, 
in the sense that they survive a time longer than their characteristic
time $\tau$, travel a distance larger than their size $A$, consistently
with results for bundles of vortex rings\cite{Alamri}. Remarkably,
vortex bundles
survive reconnections with other bundles without disintegrating,
but rather amplifying their vortex length. Finally,
the coiling of the vortex strands which is triggered by 
bundle reconnections confirms results of Holm and Kerr \cite{Kerr}
about the generation of helicity in
nearly singular vortex interactions of the Euler equation.

C.F.B. is indebted to W.F. Vinen for discussions.


\begin{thebibliography}{99}
\bibitem{proxy}
The identification of length and energy is correct under the Local
Induction Approximation \cite{Schwarz},
but is only an approximation under the Biot--Savart law.
\end{thebibliography}
\end{document}